\documentclass[twocolumn,preprintnumbers,amsmath,amssymb]{revtex4}

\usepackage{graphicx}
\usepackage{dcolumn}
\usepackage{bm}
\usepackage[final]{pdfpages}
\begin{document}
\title{Multiple nodeless superconducting gaps in optimally-doped SrTi$_{1-x}$Nb$_{x}$O$_{3}$}
\author{Xiao Lin$^{1}$,  Adrien Gourgout$^{1,2}$, German Bridoux$^{1}$\footnote{Present Address: Consejo Nacional de Investigaciones Científicas y T\'ecnicas, CONICET, Tucuman, 4000, Argentina}, Fran\c{c}ois Jomard$^{3}$, Alexandre Pourret$^{2}$,  Beno\^{\i}t Fauqu\'e$^{1}$, Dai Aoki$^{2,4}$ and Kamran Behnia$^{1}$\email{kamran.behnia@espci.fr}}
\affiliation{(1) Labotoire Physique et Etude de Mat\'{e}riaux-CNRS/ESPCI/UPMC, Paris, F-75005, France\\
(2)Institut Nanosciences et Cryog\'enie, SPSMS, CEA/UJF, Grenoble, F-38054, France\\
(3)	Groupe d'\'etude de la mati\'ere condens\'ee, F-78035 Versailles, France\\
(4) Institute for Materials Research, Tohoku University, Oarai, Ibaraki, 311-1313, Japan}

\date{September 7, 2014}

\begin{abstract}
We present the first study of thermal conductivity in  superconducting SrTi$_{1-x}$Nb$_{x}$O$_{3}$, sufficiently doped to be near its maximum critical temperature.  The bulk critical temperature, determined by the jump in specific heat, occurs at a significantly lower temperature than the resistive T$_{c}$. Thermal conductivity, dominated by the electron contribution, deviates from its normal-state magnitude at bulk T$_{c}$, following a Bardeen-Rickayzen-Tewordt (BRT) behavior, expected for thermal transport by Bogoliubov excitations. Absence of a T-linear term  at very low temperatures rules out the presence of nodal quasi-particles. On the other hand, the field dependence of thermal conductivity points to the existence of at least two distinct superconducting gaps. We conclude that optimally-doped strontium titanate is a multigap nodeless superconductor.
\end{abstract}
\maketitle

\begin{figure}\resizebox{!}{0.7\textwidth}
{\includegraphics{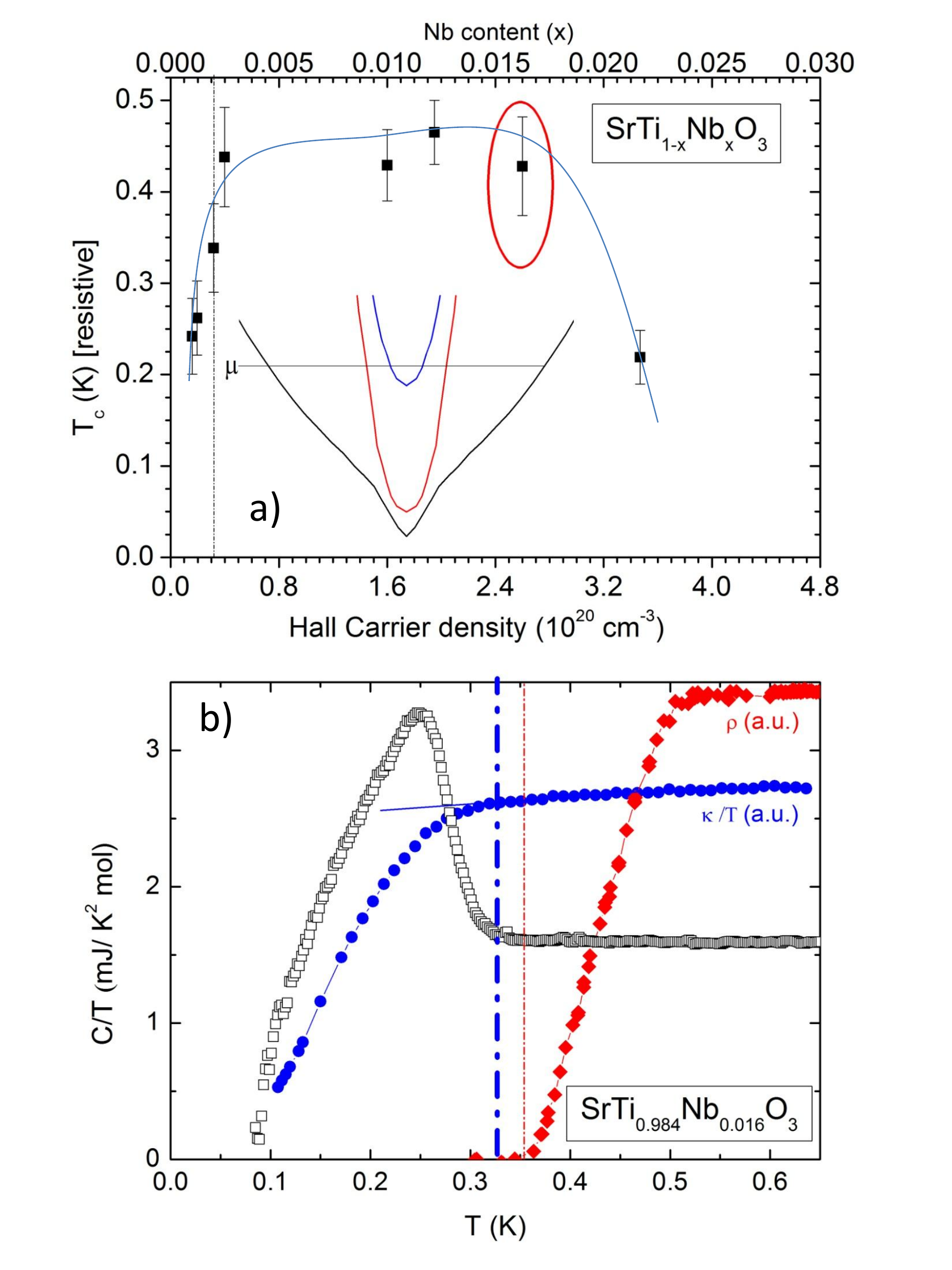}}
\caption{a) Variation of resistive critical transition in Nb-doped SrTiO$_{3}$ with carrier concentration (determined by Hall coefficient). Above the critical doping marked by a vertical line, three bands are occupied as sketched in the inset. The symbol representing the sample subject to in-depth study in this work is surrounded by a red ellipse. b) The temperature dependence of heat capacity divided by temperature (empty squares), compared with thermal conductivity divided by temperature (solid circles) and electrical resistivity (solid diamonds). The dot-dash vertical lines mark the end of the resistive and the beginning of the bulk superconducting transitions.}
\end{figure}

Discovered as early as 1964\cite{schooley1964}, the superconducting state of n-doped SrTiO$_{3}$ occupies a singular place in the history of superconductivity. Besides being the first oxide superconductor, it was one of the earliest ``semiconducting superconductors''\cite{hulm1970}, the first experimentally-detected multi-gap superconductor\cite{binnig1980} and the first case of a superconducting dome\cite{schooley1965}. Half a century after its discovery, it remains the most dilute superconductor with its Cooper pairs springing out of the tiniest Fermi surface known to undergo such a transition\cite{lin2013}.

In spite of this importance, and the intense attention devoted to the two-dimensional superconductivity discovered in SrTiO$_{3}$ heterojunctions\cite{reyren2008},  the symmetry of the superconducting order parameter of this system has remained a virgin field of exploration. Besides the early report on two distinct superconducting gaps detected by planar tunneling experiments\cite{binnig1980}, little is known about the superconducting gap and its structure. In absence of any experimental data coming from bulk probes, the existence of nodes in the superconducting gap remains an open question. Moreover, in contrast to early tunneling experiments, a recent study on superconducting interfaces did not detect multiple superconducting gaps\cite{richter2013}.

In the past two decades, thermal conductivity has emerged as a sensitive probe of nodal quasi-particles\cite{shakeripour2009}. It has been used to detect nodal and nodeless gaps of a variety of unconventional superconductors. The list includes heavy-fermion \cite{suderow1998,machida2012}, cuprate\cite{taillefer1997,nakamae2000,proust2002,proust2005}, ruthenate\cite{suzuki2002}, organic\cite{belin1997,belin1998} and iron-based\cite{tanatar2011,reid2012} superconductors. Thermal conductivity has also been used to establish the multiplicity of superconducting gaps. Beyond the emblematic case of MgB$_{2}$\cite{sologubenko2002}, several other extensively-studied superconductors, including NbSe$_{2}$\cite{boaknin2003} and CeCoIn$_{5}$\cite{tanatar2005,seyfarth2008} were identified as multi-band superconductors thanks to thermal conductivity measurements. In this paper, we present the first study of thermal conductivity in optimally-doped strontium titanate and find unambiguous evidence for the absence of nodal quasi-particles and for the multiplicity of superconducting gaps. This paves the way for the identification of the symmetry of the superconducting order parameter and the determination of the relative weight of interband and intraband pairing strength\cite{fernandes2013}.

The SrTiO$_{3}$:Nb single crystals used in this study were commercial substrates like those used in previous studies on metal-insulator transition \cite{spinelli2010}and fermiology\cite{lin2013,lin2014} in n-doped SrTiO$_{3}$. Fig.1a presents the doping dependence of resistive superconducting transition in these samples. The carrier concentration was determined by measuring the Hall coefficient and was found to be in good agreement with the expected value  according to the nominal Nb content. The latter was directly checked by Secondary Ion Beam Mass Spectroscopy (SIMS) (See the supplement).

The sample chosen for an extended study has a resistive critical temperature as high as 0.44 K (Fig. 1a). Thanks to its relatively high carrier concentration, it has a predominant electronic contribution to its specific heat and thermal conductivity. This sample was cut into two pieces. One was used for a study of heat capacity and the other for transport. Thermal conductivity, concomitant with electrical resistivity, was measured by a standard one-heater-two-thermometers set-up. Measurements were also carried out on three other samples with different carrier concentration and led to similar results (See the supplement). The focus here will be on the extended set of data obtained on one particular sample.

Fig. 1b presents the jump in specific heat caused by superconducting transition. An early measurement, limited downward to 0.3 K, detected the beginning of an upturn in specific heat\cite{ambler1966}. Our observation definitely confirms that this is a phase transition of bulk electrons. Two other pieces of information can be extracted from the specific heat data. The first is the onset temperature of bulk superconductivity. As seen in the figure, the jump starts at 0.33 K, significantly lower than the temperature at which resistivity vanishes ($\sim$ 0.35 K). Second, the magnitude of the T-linear electronic specific heat in the normal state ($\gamma \simeq 1.55$ mJ mol$^{-1}K^{-2}$) is remarkably large for a dilute metal with a carrier concentration of n$_{H}$=2.6$\times$ 10$^{20}$ cm$^{-3}$. Copper, with a carrier density 300 times larger has a $\gamma$ twice lower. The large $\gamma$ is a consequence of significant mass enhancement in doped SrTiO$_{3}$. If all electrons were in a single spherical Fermi surface, using the expression $\gamma=\frac{m^{*}k_{F}}{3}(\frac{k_{B}}{\hbar})^{2}$, one would find an effective mass of m$^{*}$=4.2 m$_{e}$. This agrees with the conclusions of the study of quantum oscillations, according to which, when the carrier concentration exceeds 2$\times$10$^{19}$ cm$^{-3}$, three bands are occupied with most carriers residing in the lowest and heaviest band with an effective mass as large as $4m_{e}$\cite{lin2014}. Thus, two distinct experimental probes converge in finding an effective mass twice as large as the band mass\cite{vandermarel2011}.

The thermal conductivity data is also shown in Fig. 1b. As seen in the figure, $\kappa/T$ does not show any detectable feature at resistive T$_{c}$, but starts to deviate from its normal-state value as soon as the jump in specific heat starts. The bulk phase transition begins at a temperature, which is 20 mK lower than the end of the resistive transition. We will come back to this feature below.

\begin{figure}
\resizebox{!}{0.35\textwidth}
{\includegraphics{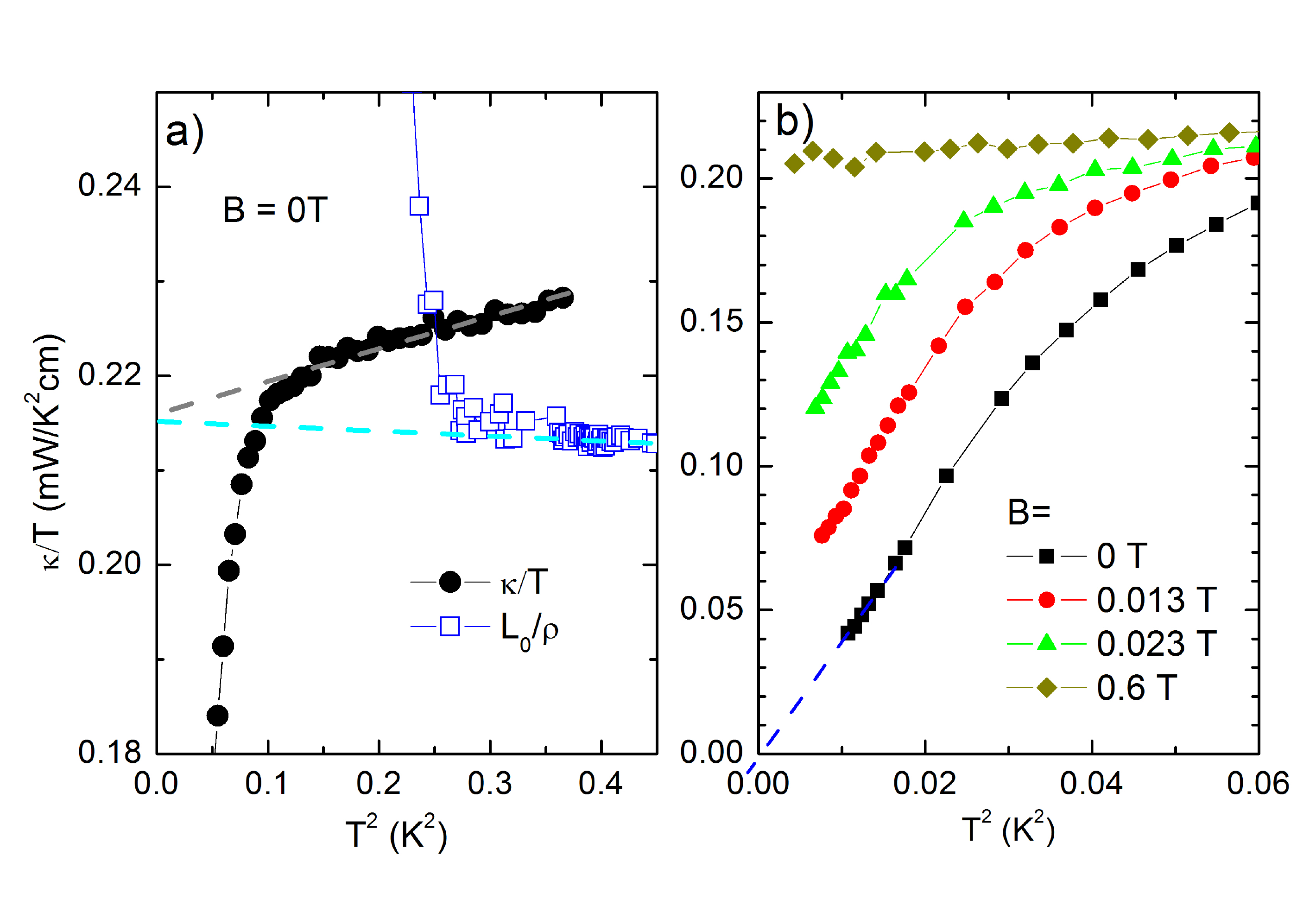}}
\caption{ a) In agreement with the Wiedemann-Franz law, the thermal conductivity divided by temperature, $\kappa/T$ (solid circles)  and Lorenz number divided by resistivity, L$_{0}$/$\rho$ (empty squares) of the normal state (T$>$T$_{c}$)  extrapolate both to the same value.  b)  $\kappa/T$, as a function of T$^{2}$ for different magnetic fields. At zero magnetic field, extrapolating $\kappa/T$ to zero-temperature leads to a vanishing intercept.}
\end{figure}

The first experimental check on the accuracy of our thermal conductivity data is the verification of the Wiedemann-Franz (WF) law. As seen in Fig. 2a, the normal-state thermal conductivity extrapolates to slightly below 0.22 mWK$^{-2}cm^{-1}$ at zero temperature. Given the residual resistivity of the sample (110 $\mu \Omega$ cm), one finds a Lorenz number close to the expected L$_{0}$=2.44 $\times$10$^{-8}$ V$^{2}$K$^{-2}$. As seen in the figure,  at the onset of resistive superconducting transition,  $\frac{L_{0}}{\rho}$ is nine-tenth of $\kappa/T$, implying an electron contribution roughly ten times the lattice component of heat transport. This simplifies the analysis and strengthen the conclusion.

Figure 2b presents the temperature dependence of thermal conductivity at different magnetic fields. At 0.6 T, superconductivity is destroyed and thermal conductivity yields a large linear term, very close to the zero-field value and the expected WF value. In the absence of magnetic field, in the superconducting state, thermal conductivity  rapidly decreases with decreasing temperature. As seen in the figure, in this case,  a linear extrapolation of  $\kappa/T$ vs. $T^{2}$ to T=0 has no detectable intercept. This result, reproduced on other samples  with different carrier concentration and mobilities (See the supplement), is the first main result of this work. In presence of nodal quasi-particles, one would expect a residual T-linear term in thermal conduction, on top of phononic T$^{3}$ term\cite{taillefer1997}. This is clearly not our case.

\begin{figure}
\resizebox{!}{0.65\textwidth}
{\includegraphics{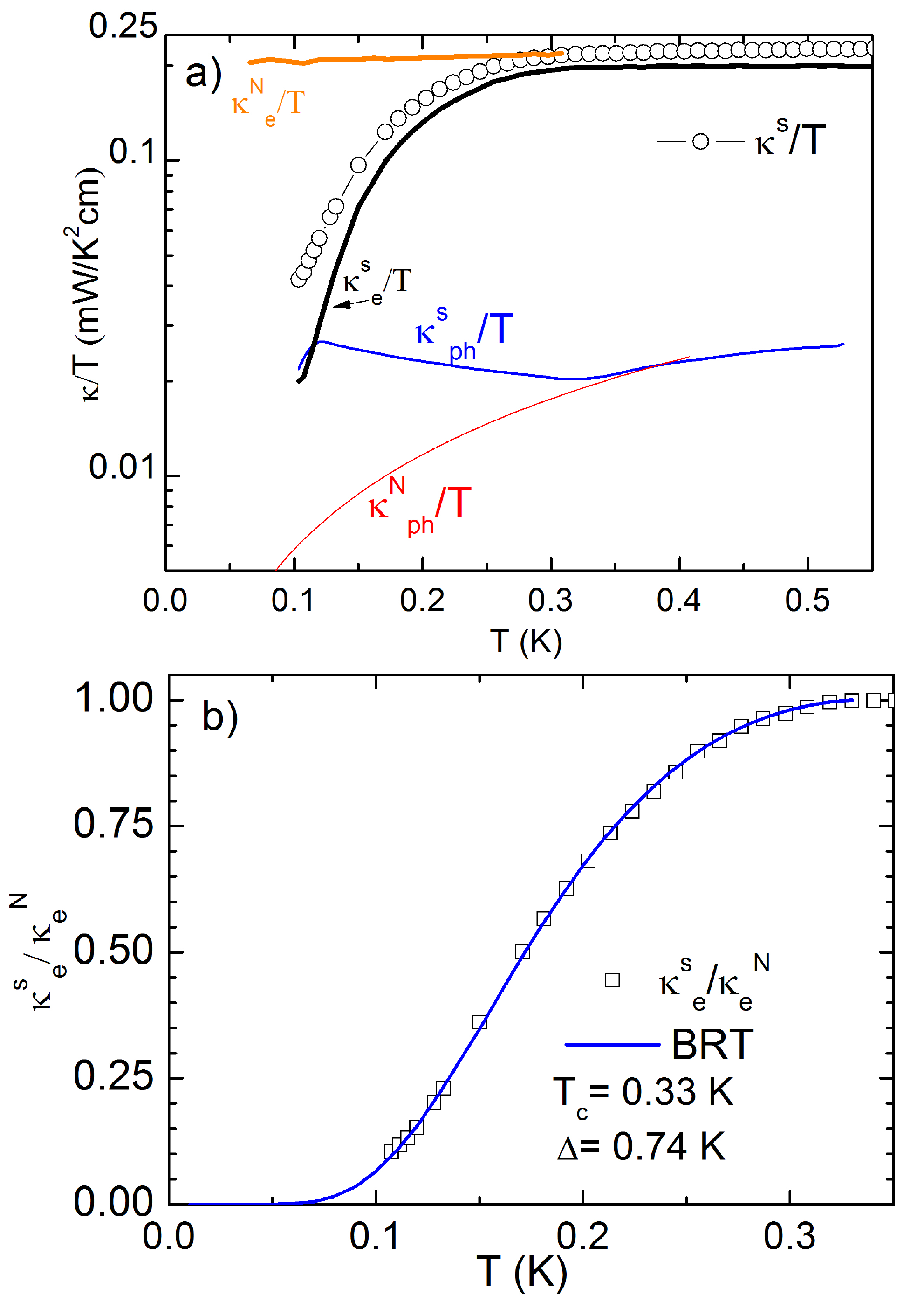}}
\caption{ a) Zero-field thermal conductivity of the superconducting state (open circles) as a function of temperature. Electronic and lattice components of thermal conductivity in normal (B=0.6T) and superconducting (B=0) states are shown as solid lines. In the normal state,  the Wiedemann-Franz law allows an unambiguous  determination of $\kappa_{ph}^{N}$ and  $\kappa_{e}^{N}$. In the superconducting state, $\kappa_{ph}^{s}$  and $\kappa_{e}^{s}$ were estimated by assuming that the phonon mean-free-path at T$_{c}$ linearly increases  to a saturated maximum of 150 $\mu$m at T=0.1K.  b) The best fit of $\kappa_{e}^{s}$ to a BRT function yielding the gap and the T$_{c}$ as parameters.}
\end{figure}

The T$^3$ term of thermal conductivity sets an upper limit to the phonon conductivity. Using the kinetic formula, $\kappa_{ph}$=1/3 C$_{ph}$v$_{s} \ell_{ph}$ and the reported values for lattice heat capacity (C$_{ph}$=$\beta$ T$^{3}$; $\beta$=JK$^{-4}$m$^{-3}$\cite{ambler1966}) and average sound velocity ($<$v$_{ph}>$=5300 ms$^{-1}$\cite{rewhald1970}), one can estimate the upper limit to the phonon mean-free-path of phonons. If $\kappa_{ph}$=bT$^{3}$ with b=3.8 mWK$^{-4}$cm$^{-1}$ one obtains a phonon mean-free-path of 0.35 mm, to be compared with sample thickness of 1 mm. Thus, even  at temperatures as low as 0.1 K, and in the absence of scattering by electrons, phonon transport may not be fully ballistic. We note that the structural transition at 105 K, in absence of strain, would create three equivalent tetragonal domains in a SrTiO$_{3}$ single crystal. The domain boundaries, which are macroscopically long \cite{buckley1999}, may play a role in setting the ultimate low-temperature mean-free-path of phonons.

In presence of B=0.6 T, when the normal state survives down to the lowest temperature, one can use the Wiedemann-Franz to separate the electronic, $\kappa_{e}^{N}$, and lattice, $\kappa_{ph}^{N}$, components of thermal conductivity. In the superconducting state, on the other hand, one cannot separate $\kappa_{e}^{s}$ and $\kappa_{ph}^{s}$ in a straightforward fashion. By vanishing in to the superconducting condensate, electrons open the road for an enhanced lattice conductivity of unknown magnitude. An upper bound to the lattice conductivity is given by the magnitude of the asymptotic T$^{3}$ thermal conductivity.  By assuming that the phonon mean-free-path linearly increases from T$_{c}$ to its maximum value at 0.1 K , with $\kappa_{ph}^{s}$ accounting for half of the total thermal conductivity, we found a $\kappa_{e}^{s}(T)$ in excellent agreement to the Bardeen, Rickayzen and Tewordt (BRT) function with plausible parameters (a T$_{c}$ of 0.33 K and a gap of 0.74 K). This function is the cornerstone of the standard theory of heat transport in a conventional superconductor\cite{bardeen1959}. The decomposition of thermal conductivity to its lattice and electronic components, as well as the fit to the BRT function are shown in the two panels of Fig. 3. Given the arbitrary assumption on $\kappa_{ph}^{s}(T)$ mentioned above, this should only be taken as a qualitative sketch. A T$_{c}$ of 0.33 K is in excellent agreement with the heat capacity data and a gap of 0.74 K (=64 $\mu$eV) is close to what was seen by tunneling measurements close to this doping range(60-80$ \mu$eV)\cite{binnig1980}. Note, however, that these would yield a $\frac{\Delta}{k_{B}T_{c}}$ ratio of 2.2, compared to the BCS value of 1.76.

\begin{figure}
\resizebox{!}{0.65\textwidth}
{\includegraphics{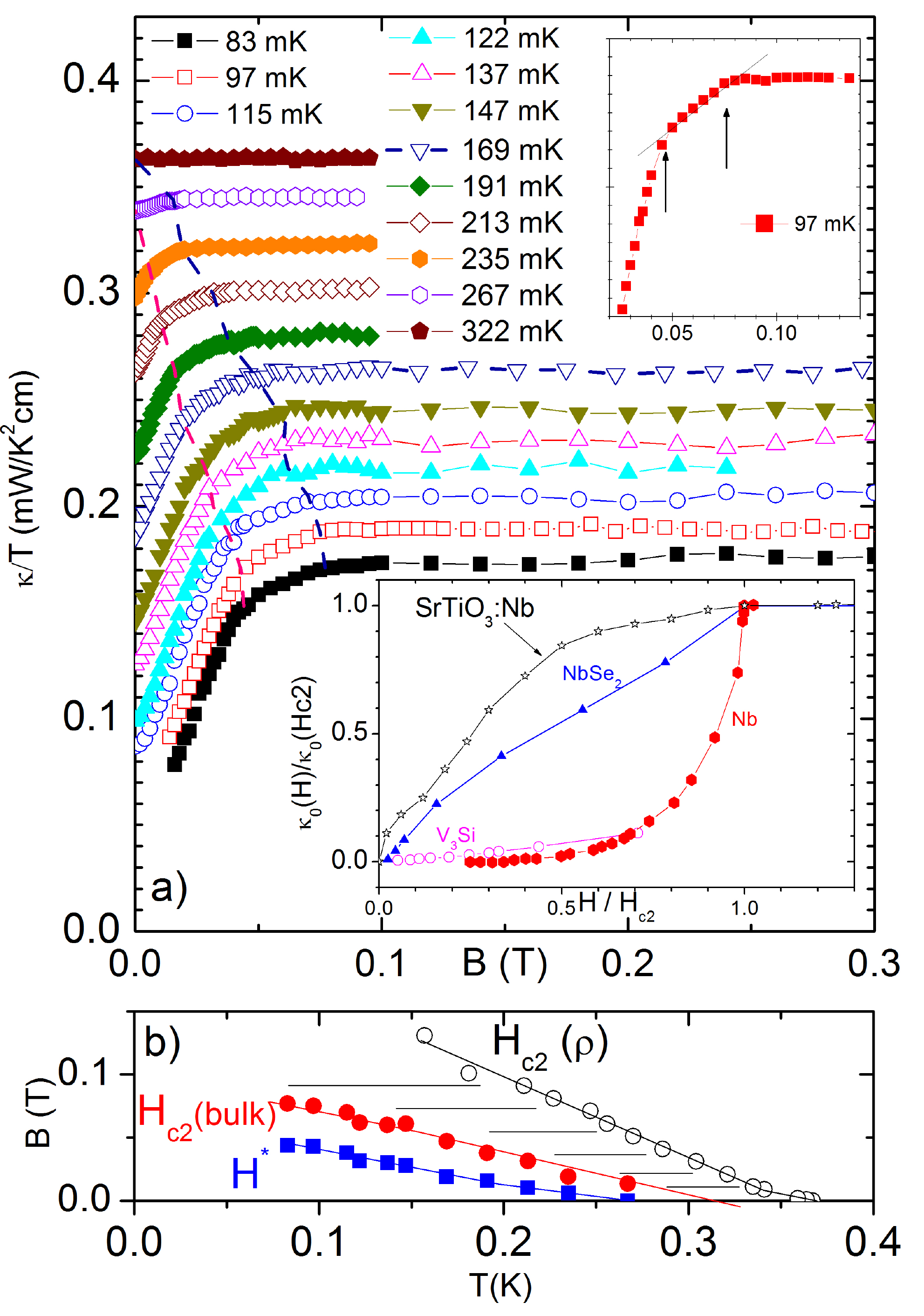}}
\caption{a) The field dependence of thermal conductivity at different temperatures reveals a shoulder at a field, H$^{*}$, below H$_{c2}$. This is more clearly see in the upper inset. The lower inset compares the field-dependence of thermal conductivity in Nb-doped SrTiO$_{3}$, a multi-gap (NbSe$_{2}$) and two single-gap (Nb and V$_{3}$Si)superconductors [See ref.\cite{boaknin2003,shakeripour2009}.  b) The two field scales extracted from $\kappa (H)$ compared to  H$_{c2}(\rho)$, the magnetic field at which resistivity vanishes. In the region filled with horizontal lines resistivity vanishes,  but bulk electrons are still normal.}
\end{figure}

 Up to here, no feature of heat transport distinguishes this superconductor from aluminum, in which the BRT function was experimentally checked decades ago\cite{satterthwaite1962}. As soon as one examines the effect of magnetic field, however, such a distinction becomes visible. As seen in Fig. 4, the application of a small magnetic field, much lower than the upper critical field, H$_{c2}$, substantially modifies the magnitude of thermal conductivity. Moreover, by sweeping the magnetic field from zero to H$_{c2}$, a shoulder is detectable in $\kappa(H)$. Both these features, were detected in multi-band superconductors\cite{sologubenko2002,boaknin2003,tanatar2005,seyfarth2008,shakeripour2009} and were interpreted as the signatures of multi-gap superconductivity.

In a nodeless single-band superconductor, the application of magnetic field, as far as the distance between vortices keep the quasi-particles trapped inside the normal cores of vortices, does not affect heat transport. On the other hand, in a multi-band superconductor, a modest magnetic field can significantly enhance thermal conductivity by closing the smaller gap. As seen in the lower inset of Fig. 4a, which compares our case with three other superconductors, this is the case of Nb-doped SrTiO$_{3}$.

Thermal conductivity becomes independent of magnetic field above a threshold magnetic field, which is 0.08 T at 0.097 K (See the upper inset of Fig. 4a). This field is the bulk upper critical field, H$^{bulk}_{c2}$. Below a second field scale, H$^{*}$, lower than H$^{bulk}_{c2}$, thermal conductivity shows a steeper field dependence. This second field scale points to the existence of an additional superconducting coherence length set by a second superconducting gap. We did not detect a third scale of magnetic filed in this three-band system. In this respect, our results are similar to those reported by Binnig \emph{et al.}\cite{binnig1980}, who detected two [and not three] distinct superconducting gaps. Two possibilities come to mind. Either two of the bands have gaps of almost identical magnitudes, or the one associated with the third band (and the corresponding field scale) are too small to be easily detectable. The initial rise of thermal conductivity by a small magnetic field, much lower than H$^{*}$ is either due to the existence of a third field scale much smaller than the other two, or a strong anisotropy of one of the two detected gaps.

As seen in Fig. 4b, the bulk upper critical field is significantly lower than the resistive upper critical field. The result is confirmed by specific heat data in presence of magnetic field (see the supplement). This brings us back to the shift observed at zero magnetic field between bulk transition temperature  and vanishing resistivity. In a portion of the (B,T) plane, bulk electrons are still normal, but the system shows zero resistivity. This calls for an explanation. Invoking sample inhomogeneity does not provide an answer. In a wide doping range, the critical temperature does not show a strong dependence on doping. Moreover, the observation of quantum oscillations with a well-defined frequency corresponding to the density of bulk carriers estimated from Hall effect puts an upper limit to any macroscopic inhomogeneity.

One place for superconductivity to survive when bulk electrons are normal are boundaries between tetragonal domains. If the critical temperature  happens to be higher in these twin boundaries than in the bulk, one can observe a vanishing resistivity at a temperature well above the bulk critical temperature. Recent near-field studies on the STO interfaces have detected enhanced electrical conductivity along twin boundaries \cite{kalisky2013,honig2013}, providing plausibility to this speculation.

In summary, we find that optimally-doped SrTiO$_{3}$:Nb is a multi-gap superconductor and none of its gaps has nodes. These are new pieces in this puzzle of exceptionally dilute superconductivity, for which several exotic pairing mechanisms (such as a phonon soft mode \cite{appel1969}, plasmons\cite{takada1980} or ferroelectric quantum criticality\cite{rowley2014}) have been proposed. In contrast to all other known cases of multi-band superconductivity, one can here tune the Fermi surface (in sheer size as well as the number of its components) across several orders of doping concentration. This provides new experimental opportunities and stronger constraints for theory.

This work is supported by Agence Nationale de la Recherche as a part of QUANTHERM and SUPERFIELD projects.

\newpage

\appendix
\section{Specific heat in presence of magnetic field}

Fig. S1 presents the specific heat data in presence of magnetic field. As seen in the figure,  the jump associated with the superconducting transition shifts to lower temperatures  with increasing magnetic field.  Applying a field as small as 0.06 T, lowers the temperature of the jump to around 0.2 K. Thus, the data confirms that bulk superconductivity is restricted to a limited region in the (field, temperature) plane significantly lower than what is indicated by resistivity.

\begin{figure}
\renewcommand{\thefigure}{S1}
\resizebox{!}{0.35\textwidth}
{\includegraphics{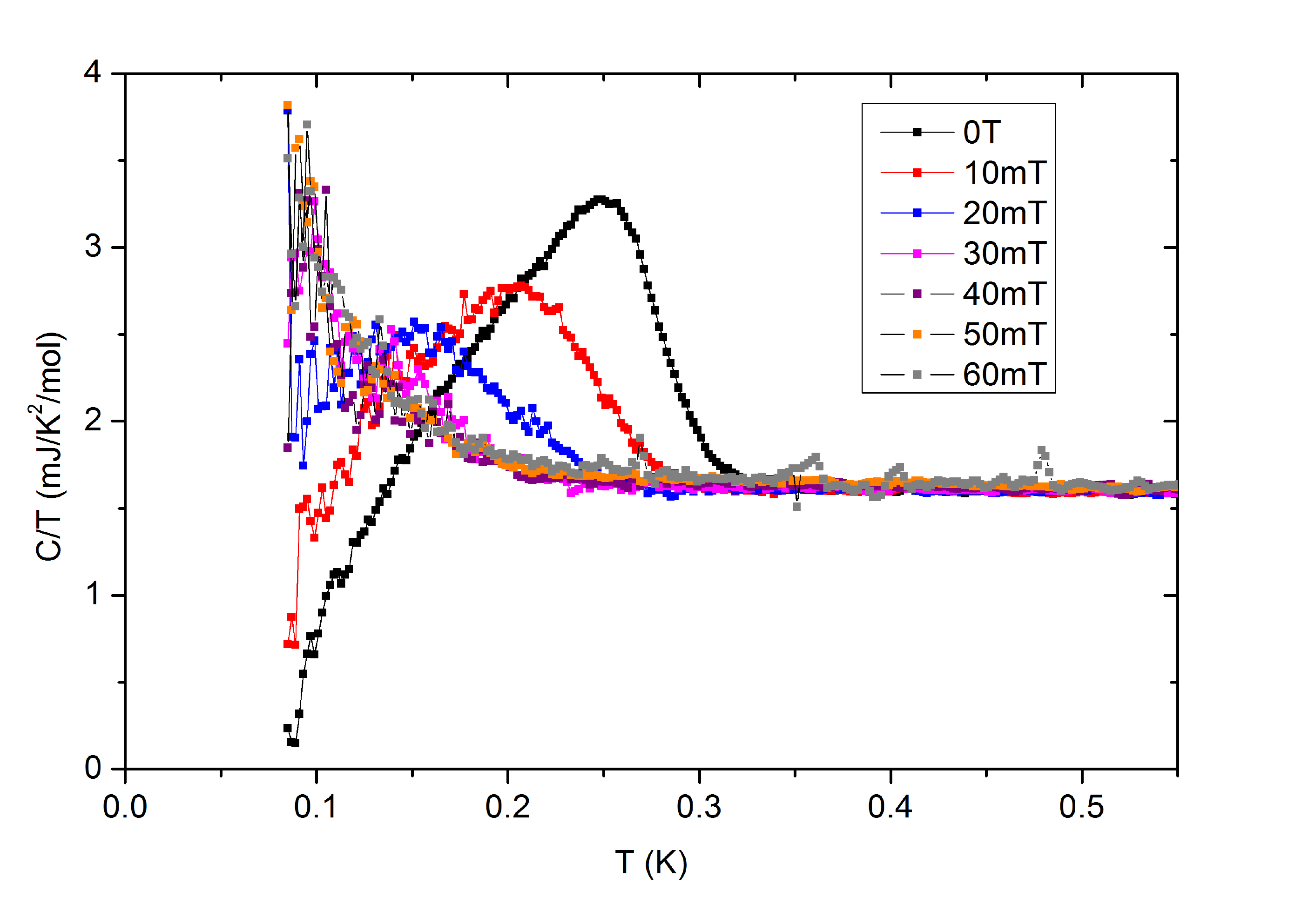}}
\caption{ Specific heat of SrTi$_{0.884}$Nb$_{0.016}$O$_{3}$ in presence of magnetic field. }
\end{figure}

The signal-to-noise ratio in the specific heat data does not allow the resolution a second field scale in addition to upper critical field. Moreover,  at low temperature the signal is contaminated by the presence of a Schottky anomaly and uncertainty on the contribution by the addenda.

\section{Low-temperature thermal conductivity in different samples}

\begin{figure}
\renewcommand{\thefigure}{S2}
\resizebox{!}{0.35\textwidth}
{\includegraphics{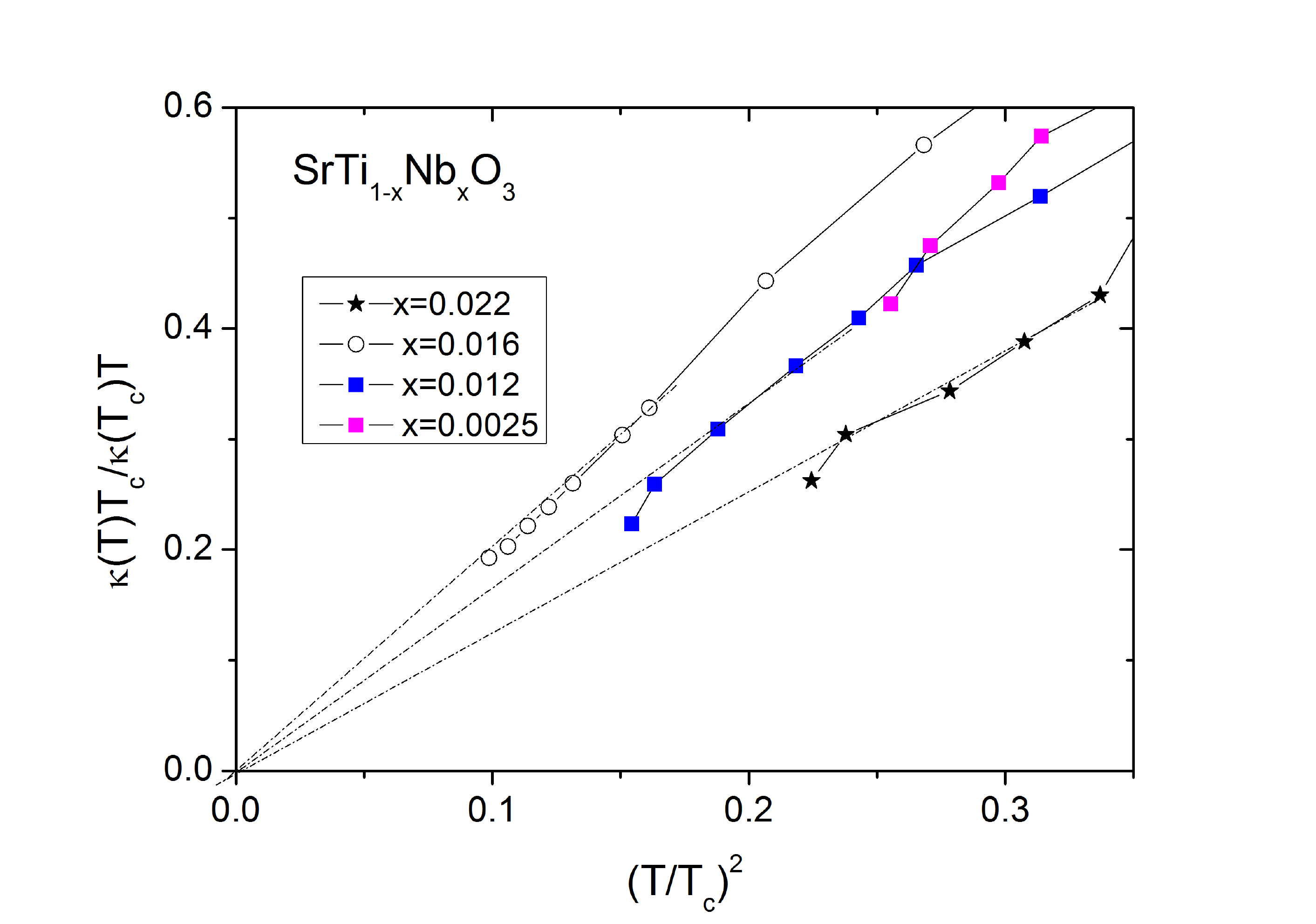}}
\caption{ Thermal conductivity in four different samples of SrTi$_{1-x}$Nb$_{x}$O$_{3}$ with different Nb content and different mobilities. Plotting the normalized $\kappa$/T as a function of normalized temperature, to compare the data in different samples, one can see that in none of the four samples present a detectable finite intercept. }
\end{figure}

The paper focused on the results obtained on a SrTi$_{1-x}$Nb$_{x}$O$_{3}$ single crystal with a carrier density of $2.5\times 10^{20}$ cm$^{-3}$ corresponding to a Nb content of x=0.016 . In addition to this sample, we measured thermal conductivity of three other SrTi$_{1-x}$Nb$_{x}$O$_{3}$ with carrier densities ranging from $4\times10^{19}$ to $3.5 \times 10^{20}$. Fig. S3  compares the thermal conductivity of all these samples at very low temperatures. As seen in the figure, in none of them can detect a finite intercept for $\kappa/T$ in the zero-temperature limit. We conclude that a linear term in the thermal conductivity  is absent or undetectable in SrTi$_{1-x}$Nb$_{x}$O$_{3}$ for $0.0025<x<0.022$.

\begin{figure}
\renewcommand{\thefigure}{S3}
\resizebox{!}{0.4\textwidth}
{\includegraphics{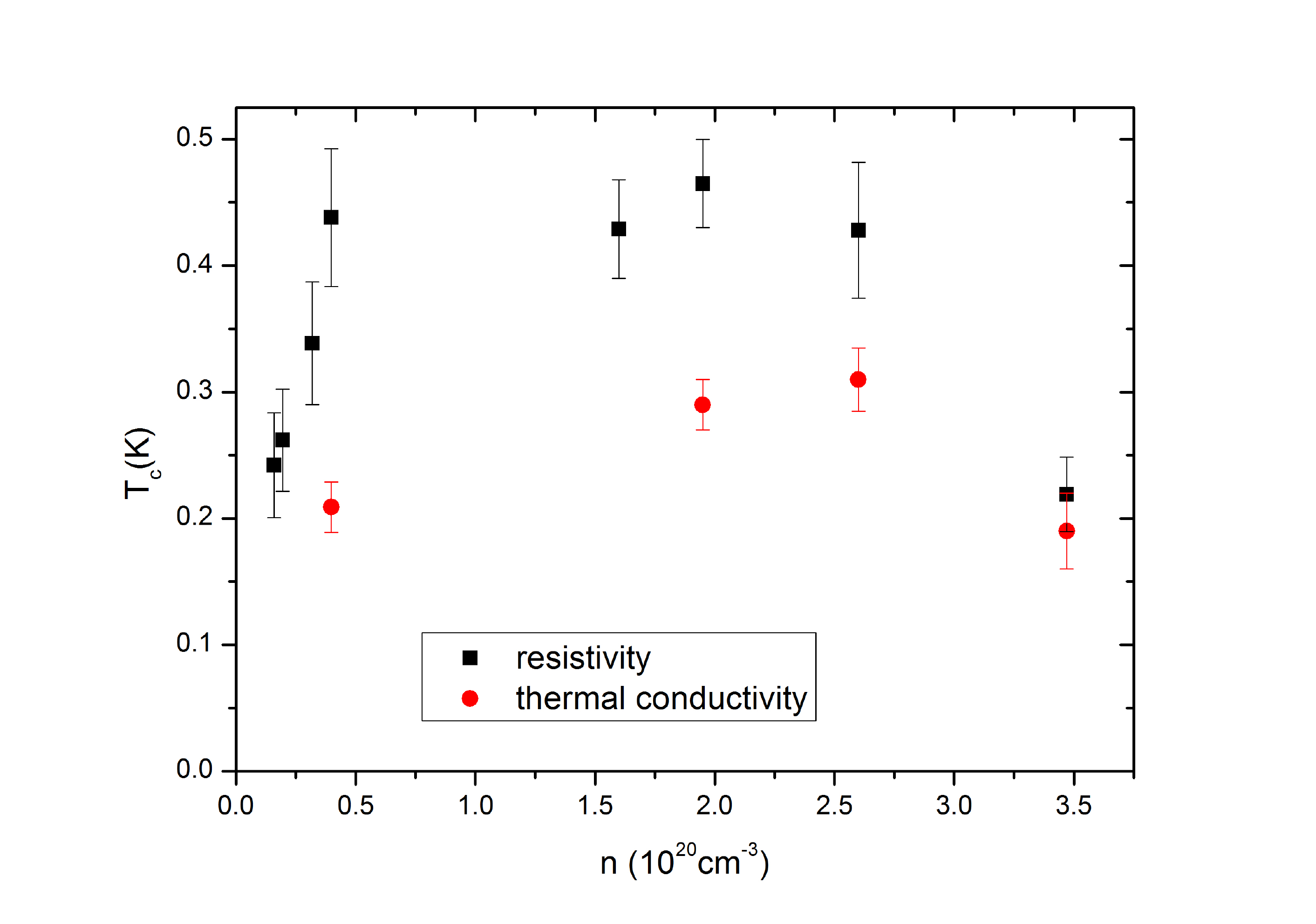}}
\caption{  The variation of the critical temperature with doping according to resistivity and thermal conductivity. In all samples, the onset of the drop in thermal conductivity indicative of the bulk superconducting transition temperature occurs below the resistive critical temperature. }
\end{figure}

In these three samples, like the one which was the focus of the study, we found that the drop in $\kappa$/T occurs at a temperature, which is significantly lower than the temperature at which resisitivity vanishes (See fig. S4).  Together with the sample shown in the main paper, three Nb doped SrTiO$_3$ around optimal doping clearly show difference on temperature between the bulk and resistive superconducting transitions. In this doping window, the critical temperature does not show a strong doping dependence. Therefore, it is unlikely that sample inhomogeneity is the origin of this discrepancy.

\section{cross-checking niobium content with SIMS}

\begin{figure}
\renewcommand{\thefigure}{S4}
\resizebox{!}{0.35\textwidth}
{\includegraphics{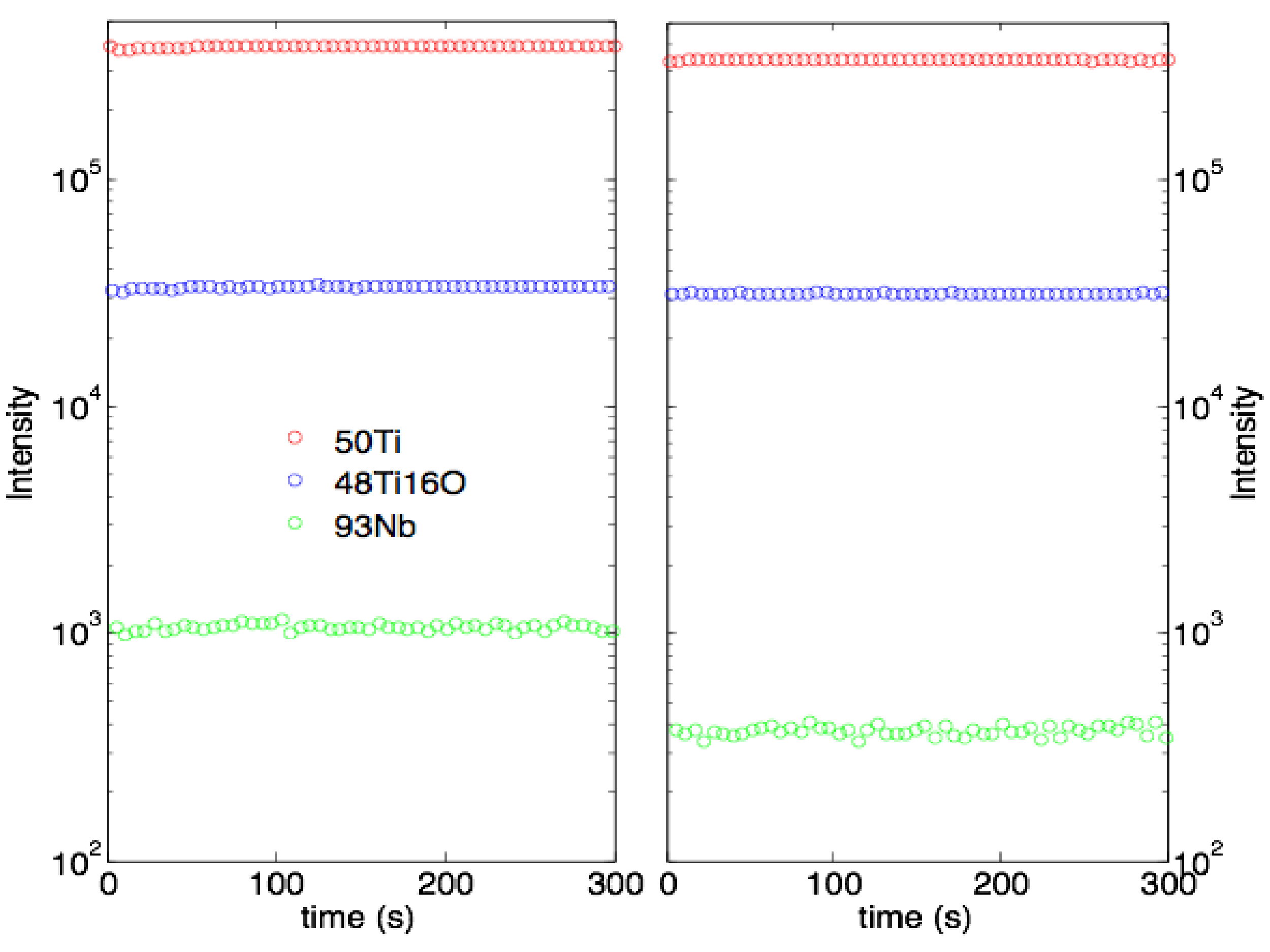}}
\caption{  SIMS profiles in two SrTi$_{1-x}$Nb$_{x}$O$_{3}$ single crystals. Left: x=0.00056; Right: x=0.0002}
\end{figure}

The nominal Nb concentration in several Nb-doped SrTiO$_{3}$ samples was cross-checked using the SIMS (Secondary Ion Mass Spectrometry). This analysis technique allows to detect very low concentrations of dopants and to measure their distribution over a depth of few microns. The measurements were performed using an IMS Cameca 7f and an oxygen primary beam.

Fig. S5 shows a typical profile of three atomic masses M=50, 80 and 93 in two samples. These correspond respectively to Ti atoms, the ionic complex TiO$_{2}$ and Nb atoms. In order to decrease the ionic molecular formation, we used a polarized source at -150V. The intensities are constant as a function of time, which indicates that the concentration of each element does not change as one probes deeper  perpendicular to the surface. As seen in the figure, the intensity of the two reference masses (M=50[Ti] and 80[48Ti16O2]) are comparable, but the one corresponding to the Nb mass is different.

By comparing the intensity for the Nb mass in the two different samples one finds a very good agreement with the nominal content. The first sample (x=0.00056) is expected to have 2.8 times more Nb than the second sample(x=0.0002). The ratio of the M=93[Nb] to M=50[Ti] signals is 2.54 times larger in the first sample compared to the second one. Comparing the ratio of the M=93[Nb] to  80[48Ti16O2]) signals, one finds that it is 2.69 times larger in the first sample than in the second sample. These numbers are in reasonable agreement with the nominal content. They confirm that SIMS is a powerful probe of relative (and not necessarily absolute) dopant content, resolving differences as small as a fraction of 10$^{-4}$ between two SrTi$_{1-x}$Nb$_{x}$O$_{3}$ samples.

\begin{thebibliography}{}
\bibitem{schooley1964}  J. F. Schooley, W. R. Hosler, and M. L. Cohen, Phys. Rev. Lett. \textbf{12}, 474 (1964)
\bibitem{hulm1970} J. K. Hulm, D. Ashkin, D. W. Deis, and C. K. Jones, Prog. in Low Temp. Phys. VI, 205 (1970)
\bibitem{binnig1980} G. Binnig, A. Baratoff, H. E. Hoenig, and  J. G. Bednorz, Phys. Rev. Lett. \textbf{45}, 1352 (1980)
\bibitem{schooley1965} J. F. Schooley\emph{ et al.}, Phys. Rev. Lett. \textbf{14}, 305 (1965)
\bibitem{lin2013} X. Lin, Z. Zhu, B. Fauqu\'{e}, and K. Behnia, Phys. Rev. X \textbf{3}, 021002 (2013)
\bibitem{reyren2008} N. Reyren \emph{et al.}, Science \textbf{317}, 1196 (2007)
\bibitem{richter2013} C. Richter \emph{et al.},  Nature \textbf{502}, 528 (2013)
\bibitem{shakeripour2009} For a review see: H. Shakeripour, C. Petrovic, and L. Taillefer, New J. Phys. \textbf{11}, 055065 (2009)
\bibitem{suderow1998} H. Suderow, J. P. Brison, A. Huxley, and J. Flouquet, Phys. Rev. Lett. \textbf{80}, 165 (1998)
\bibitem{machida2012} Y. Machida \emph{et al.}, Phys. Rev. Lett. \textbf{108}, 157002  (2012)
\bibitem{taillefer1997} L. Taillefer \emph{et al.}, Phys. Rev. Lett. \textbf{79}, 483 (1997)
\bibitem{nakamae2000} S. Nakamae \emph{et al.}, Phys. Rev. B \textbf{63}, 184509 (2000)
\bibitem{proust2002}C. Proust \emph{et al.} Phys. Rev. Lett. \textbf{89}, 147003 (2002)
\bibitem{proust2005} C. Proust \emph{et al.}, Phys. Rev. B \textbf{72}, 214511 (2005)
\bibitem{suzuki2002}M. Suzuki \emph{et al.}, Phys. Rev. Lett. \textbf{88}, 227004(2002)
\bibitem{belin1997} S. Belin, and K. Behnia, Phys. Rev. Lett. \textbf{79}, 2125 (1997)
\bibitem{belin1998} S. Belin, K. Behnia, and A. Deluzet, Phys. Rev. Lett. \textbf{81}, 4728 (1998)
\bibitem{tanatar2011} M. A. Tanatar \emph{et al.} , Phys. Rev. B \textbf{84}, 054507 (2011)
\bibitem{reid2012} J.-Ph. Reid \emph{et al.}, Phys. Rev. Lett. \textbf{109}, 087001 (2012)
\bibitem{sologubenko2002} A. V. Sologubenko \emph{et al.}, Phys. Rev. B \textbf{66}, 014504 (2002)
\bibitem{boaknin2003} E. Boaknin \emph{ et al.}, Phys. Rev. Lett. \textbf{90}, 117003 (2003)
\bibitem{tanatar2005} M. A. Tanatar \emph{et al.}, Phys. Rev. Lett. \textbf{95}, 067002 (2005)
\bibitem{seyfarth2008}G. Seyfarth \emph{et al.}, Phys. Rev. Lett. \textbf{101}, 046401 (2008)
\bibitem{fernandes2013} R. M. Fernandes,  J. T. Haraldsen, P. W\"{o}lfle, and A. V. Balatsky, Phys. Rev. B \textbf{87}, 014510 (2013)
\bibitem{spinelli2010} A. Spinelli \emph{et al.}, Phys. Rev. B \textbf{81}, 155110 (2010)
\bibitem{lin2014} X. Lin \emph{et al.}, Phys. Rev. Lett. \textbf{112}, 207002 (2014)
\bibitem{ambler1966} E. Ambler, J. H. Colwell, W. R. Hosler, and J. F. Schooley, Phys. Rev. \textbf{148}, 280 (1966)
\bibitem{vandermarel2011}D. van der Marel, J. L. M. van Mechelen, and I. I. Mazin, Phys. Rev. B \textbf{84}, 205111 (2011)
\bibitem{rewhald1970} W. Rewhald, Solid State Communications \textbf{8}, 607 (1970)
\bibitem{buckley1999} A. Buckley, J. P. Rivera, and E. K. H. Salje, J. Appl. Phys. \textbf{86}, 1653 (1999)
\bibitem{bardeen1959} J. Bardeen, G. Rickayzen, and L. Tewordt, Phys. Rev. \textbf{113}, 982 (1959)
\bibitem{satterthwaite1962} C. B. Satterthwaite, Phys. Rev. \textbf{125}, 873 (1962)
\bibitem{kalisky2013} B. Kalisky \emph{et al.}, Nature Mater. \textbf{12}, 1091(2013)
\bibitem{honig2013} M. Honig \emph{et al.} Nature Mater. \textbf{12}, 1112 (2013)
\bibitem{appel1969} J. Appel, Phys. Rev. \textbf{180}, 508 (1969)
\bibitem{takada1980} Y. Takada,  J. Phys. Soc. Jpn. \textbf{49} 1267 (1980)
\bibitem{rowley2014}S. E. Rowley \emph{et al.}, Nature Phys. \textbf{10}, 367 (2014)
\end{thebibliography}
\end{document}